\documentclass[11pt]{article}
\usepackage{amssymb}
\usepackage{pgf}
\usepackage{tikz}
\usepackage{amsmath}
\usepackage{framed}
\usepackage{rotating}
\usepackage{amsthm}
\usepackage{amsfonts}
\usepackage{proof}

\usepackage{geometry}
\usetikzlibrary{arrows,automata}
\usetikzlibrary{positioning}

\theoremstyle{definition}
\newtheorem{definition}{Definition}[section]

\theoremstyle{remark}

\usepackage{graphicx} 

\usepackage{amssymb}
\usepackage{graphicx}
\usepackage{color}

\usepackage{booktabs} 
\usepackage{array} 
\usepackage{paralist} 
\usepackage{verbatim} 
\usepackage{subfig} 
\usepackage{mathtools}

\usepackage{fancyhdr} 
\usepackage{url}
\urldef{\mailsa}\path|{emanuela.merelli,matteo.rucco, luca.tesei}@unicam.it|

\usepackage{sectsty}
\allsectionsfont{\sffamily\mdseries\upshape} 

\usepackage[nottoc,notlof,notlot]{tocbibind} 
\usepackage[titles,subfigure]{tocloft} 

\newcommand{\goes}[1]{\xrightarrow[]{#1}}

\title{TOPDRIM deliverable 3.2 \\ September 2014\bigskip \\
Topological characterization of S[B] systems: From data to models of complexity
}

\usepackage{authblk}

\author[1]{Emanuela Merelli\thanks{emanuela.merelli@unicam.it}}
\author[1]{Matteo Rucco\thanks{matteo.rucco@unicam.it}}
\author[2]{Peter Sloot\thanks{p.m.a.sloot@uva.nl}}
\author[1]{Luca Tesei\thanks{luca.tesei@unicam.it}}
\affil[1]{School of Science and Technology, University of Camerino, Italy}
\affil[2]{University of Amsterdam, The Netherlands. Complexity Institute, NTU, Singapore. ITMO University, St. Petersburg, Russian Federation}

\date{} 

\begin{document}
\maketitle

\section*{Abstract}

In this paper we propose a methodology for deriving a model of a complex system by exploiting the information extracted from Topological Data Analysis. Central to our approach is the $S[B]$ paradigm in which a complex system is represented by a two-level model. One level, the structural $S$ one, is derived using the newly introduced \textit{quantitative} concept of Persistent Entropy. The other level, the behavioral $B$ one, is characterized by a network of interacting computational agents described by a Higher Dimensional Automaton. The methodology yields also a representation of the evolution of the derived two-level model as a Persistent Entropy Automaton. The presented methodology is applied to a real case study, the Idiotypic Network of the mammal immune system.  

\subsubsection*{Keywords: Topological Data Analysis, Persistent Entropy, Transition Systems, Higher Dimensional Automata, Immune System, Complex Systems, Computational Agents.}

\section{Introduction}
\label{sec:intro} 

Complex systems are typically characterized by a finite, typically large, number of non-identical interacting entities that often are complex systems themselves, with strategies and autonomous behaviors. Complex system science is a relatively young research area and it is raising interest among many researchers mainly thanks to the emerging of new techniques in several fields such as physics, mathematics, data analysis and computer science \cite{stein2014nature, jankowski2014practical}. 

Complex systems are analyzed using mainly two different approaches: the one that provides a global and abstract description by systems of differential equations~\cite{boccara2004modeling}, the other that focuses on the interacting components of a complex systems, which are generally simulated by an \textit{agent based model and simulation framework}~\cite{holland1992complex}. 
 
In terms of data volume a complex system can be associated to a big dataset and in order to extract from it useful information new techniques have been recently introduced, most of them in the area of \textit{Topological Data Analysis} (TDA). TDA is largely used for the analysis of complex systems. Ibekwe et al. in~\cite{ibekwe2014topological} used TDA for reconstructing the relationship structure of E. coli O157 and the authors also proved that the \textit{non-O157} is in 32 soils (16 organic and 16 conventionally managed soils). TDA was also used by De Silva in~\cite{de2007coverage} for the analysis of sensor networks and it was successfully applied to the study of viral evolution in biological complex systems~\cite{chan2013topology}. In~\cite{petri2014homological} Petri et al.\ used a homological approach for studying the characteristics of functional brain networks at the mesoscopic level.

TDA is a new discipline inspired by homology theory. Informally, homology is a machinery for counting the number of $n$-\textit{dimensional holes} in a topological space. A topological space can be formed by a collection of topological objects: the simplices (vertex, line segments, triangles, tetrahedra, and so on). A collection of simplices forms a simplicial complex~\cite{munkres1984elements}. Simplicial complexes can be built in several ways, e.g., by using the Vietoris-Rips complexes, Witness complexes, neighborhood complexes, clique complexes~\cite{carlsson2009topology, petri2013topological, binchi2014jholes}. Simplicial complexes can be studied via \textit{persistent homology}~\cite{edelsbrunner2008persistent, edelsbrunner2010computational} which computes a parametrized version of \textbf{Betti numbers}, represented by \textit{Betti barcodes}. A barcode is a collection of line segments (or points in case of persistent diagrams~\cite{carlsson2004persistence}) where each line is equipped with two pieces of information: the spanning life time \textit{intervals} of the homology class associated to the line and the \textit{generators} of the homology class. An interval, written as $[i; j]$, means that the associated simplex forms a homology class at time $t=i$ and it survives until time $t=j$. If $j=\infty$, then the homology class is \textit{persistent} and the interval is written $[i; +\infty)$; otherwise it is classified as \textit{topological noise}. The generators are the set of simplices involved in the homology class.

In this work we propose a methodology for modeling complex systems that is based on TDA. Our contribution has been inspired by the \textit{\textbf{S[B]}} paradigm~\cite{merelli2013non,merelli2014topology,MPT15}. Even if $S[B]$ has been thought as a general theoretical framework for modeling complex \textit{software} systems, we argue that it can be used to model all kinds of complex systems (biological, physical, chemical, and so on) with the following features:

\begin{itemize}
\item the system is formed by huge amount of interacting components, generally belonging to different classes and with different behaviors. In this kind of systems it is usually interesting to observe emerging global behaviors~\cite{holland1992complex};
\item the system can be seen as the composition of two levels: a level characterized by the interacting components and a  level containing the set of constraints and/or initial conditions that guide the system behaviour;
\item the set of constraints can evolve over time. The time evolution of the set of constraints is well-known as \textit{internal memory} or \textit{adaptability}~\cite{fayad1996aspects};
\item all entities can simultaneously execute several actions in response to internal and/or external stimuli. The system behavior is like to that of a concurrent and distributed system studied in \textit{concurrency} theory~\cite{milner1989communication,hoare1985communicating,bergstra1985algebra}
\item each interaction can be characterized by a parameter that we call ``proximity''.
\end{itemize}

\begin{figure}
\center
\includegraphics[width=\textwidth]{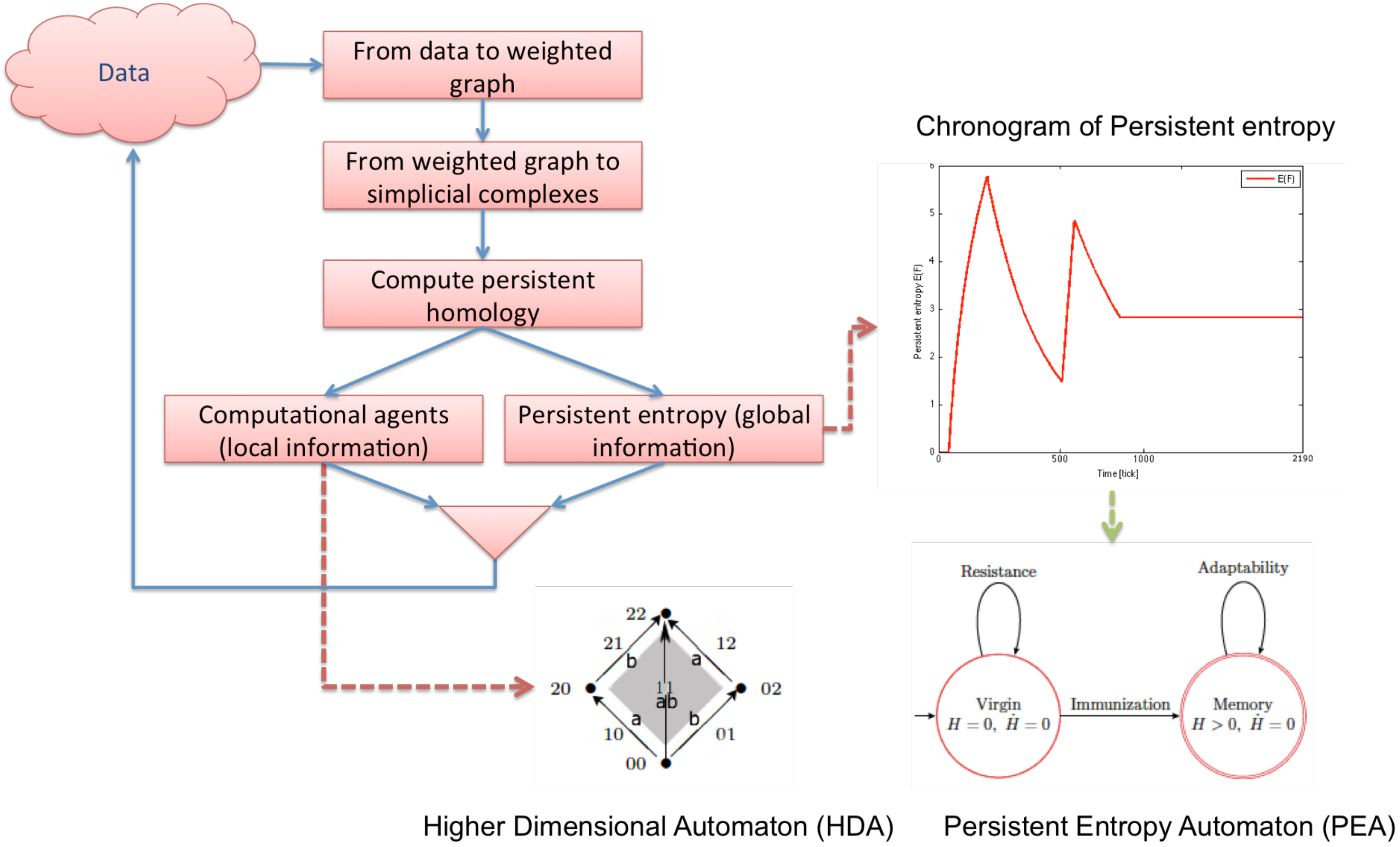}
\caption{Graphical representation of our methodology. All the details are explained in the text, especially Persistent Entropy in Sect 3.4}
\label{fig:diagram}
\end{figure}

Figure \ref{fig:diagram} graphically shows our methodology. It is an iterative process, where each iteration corresponds to a data observation taken from the complex system under study. In each iteration, data are first converted to a weighted graph in order to apply the subsequent steps. The graph is taken as the scaffold of a topological space from which a representation with simplicial complexes is derived. Then, persistent homology is computed in order to trace the topological invariants (Betti numbers) of the underlying topological space and the generators of the homological classes. At this point, two kinds of information are derived. On the one hand, the generators are taken as the more relevant computational agents among which local interactions in the complex system occur. These agents are represented by a Higher Dimensional Automaton~\cite{pratt1991modeling} and, within the $S[B]$ paradigm, they constitute an instantaneous description of the \textit{\textbf{behavioral level B}}. On the other hand, the global dynamics of the complex system is described by a newly introduced entropic measure that we call \emph{Persistent Entropy}. Its value for the current data is plotted on a chronogram. From the plot of persistent entropy a model of the global evolution of the complex system can be derived in the form of a \emph{Persistent Entropy Automaton} that is a state machine whose transitions are constrained by the observed values of the persistent entropy and its first numerical derivative. Such an automaton, within the $S[B]$ paradigm, constitutes the \textit{\textbf{structural level S}}.

To validate our methodology we consider, as case study, the \textit{Idiotypic Network} (IN) of 
the \textit{mammal immune system}~\cite{jerne1974towards}. This system has been largely studied because it exhibits all the features typical of a complex adaptive system (CAS)~\cite{holland1992complex,dasgupta1999artificial} listed in the following.
\begin{itemize}
\item Distributed Control: Immune System (IS) is not centrally controlled. Detection and response can be executed locally and immediately without the need of communication of billions of immune molecules and cells with a central organ.
\item Connectivity: IS is suited with inter-relationship, inter-action and inter-connectivity of the elements within a system and between a system and its environment.
\item Co-evolution: the behavior of antibodies evolves depending on the behaviors of other antibodies; they can play a dual role.
\item Sensitive Dependence on Initial Conditions: the IS reacts depending on the initial condition, e.g., the antigen volume
\item Emergent-order: IS has both \textit{self-regulation} and \textit{self-protection} mechanisms.
\item Learning and memory: IS is able to learn through its interaction with the environment.
\end{itemize}

Several models for the IN are available. In particular, we consider the simplified description modeled by Parisi~\cite{parisi1990simple} which is also the base for the agent-based simulator \textit{C-ImmSim}~\cite{bernaschi2001design}. More details about the model will be given in Section~\ref{sec:introcs}. In order to conduct a controlled experiment and facilitate the task of validation, in Section~\ref{sec:casestudy} we take as data observation of the complex system produced by the simulator, not real data. In this way we can check more easily the correctness of our methodology. 

The paper is organized as follows. Section~\ref{sec:introcs} introduces the case study of the IN. In Section~\ref{sec:methodology} we describe our methodology step by step in order to extract from data a two-level model according to the $S[B]$ paradigm. In Section~\ref{sec:casestudy} we apply the methodology to the case study of Idiotypic Network and finally Section~\ref{sec:conclusions} provides concluding remarks and open issues to be faced in future works.

\section{Case Study}
\label{sec:introcs}

In this paper we illustrate the steps of the proposed methodology applying it to a case study in biological Immune System (IS).
Historically, the first theory is known as \textit{Immune Network theory} (IN) by Nielse Jerne~\cite{jerne1974towards}. Jerne theorized that the immune system can be thought as a regulated network of antibodies and anti-antibodies, called ``idiotypic network''. The network works even in the absence of antigens because the antibodies can be recognized as foreign cells, some of them previously stimulated by antigens. Let us briefly recall the main mechanisms described by the model. When an antigen is presented to the organism, the IS reacts following two ways: innate immunity and adaptive (or acquired) immunity.
Innate immune defenses are non-specific, meaning these systems respond to pathogens in a generic way. The innate immune system is the dominant system of host defense in most organisms, it involves the eithelial barriers, the phagocytes, denditric cells, plasma proteins and NK cells.  The adaptive immune system, on the other hand, is called into action against pathogens that are able to evade or overcome innate immune defenses. The first reaction of the immune system is governed by the mechanism of the innate immunity and its lifespan is up to 12 hours, after that the possible presence of antigens trigger the adpative response.   Components of the adaptive immune system are mainly B-cells, antibodies, naive T cells and Effector T cells. When activated, these components ``adapt'' to the presence of infectious agents by activating, proliferating, and creating potent mechanisms for neutralizing or eliminating the antigens. The lifespan of the adaptive response depends by the type of infection. In this work we are interested to study the behavior of the antibodies involved during the adaptive response and we report a general example of this scenario. Suppose an antigen is recognized by B cells, which secrete antibodies, say $Ab_1$. $Ab_1$ themselves are then recognized as anti-antibodies by ``anti-idiotypic'' B cells, which secrete other antibodies, say $Ab_2$. Thus, further interactions can lead to $Ab_3$ antibodies that recognize $Ab_2$ and so on. In an idiotypic network, there is no intrinsic difference between an antigen and an antibody. Moreover, any node of the network can bind to and be bound to any other antigen or antibody. This phenomenon is known as \textit{idiotypic cascade}. During the \textit{onto-genesis} phase the IS learns which antibodies should not be produced and the system remembers these decisions for its entire life. This phenomenon is called \textit{immunological memory}. Important properties of this system, including memory, are then properties of the network of cells as a whole, rather than of the individual cells~\cite{hoffmann1975theory}.  This phenomenological description has been formalized by Parisi~\cite{parisi1990simple}. Parisi derived a simplified model for describing the dynamics of a functional network of antibodies in absence of any driving force of external antigens, namely when the concentrations of antibodies are time independent, e.g. during the immune memory state. 
The dynamics $h_i$ of each antibody at the equilibrium  is simply described in Equation~\ref{eq:dynamics}:
\begin{equation}
\label{eq:dynamics}
h_i=S + \Sigma_{k=1,n}J_{i,k}c_k
\end{equation}
Where $J_{i,k}$ $(J_{i,i} = 0; J_{k,i} = J_{i,k})$ represents the influence of antibody $k$ on antibody $i$. If $J_{k,i}$ is positive, antibody $k$ triggers the production of antibody $i$, whereas if $J_{k,i}$ is negative, antibody $k$ suppresses the production of antibody $i$. $|J_{k,i}|$ is a measure of the efficiency of the control of antibody $k$ on antibody $i$. The $J_{i,k}$ are distributed in the interval $[-1; +1]$. S is a threshold parameter, that regulates the dynamics when the couplings$J_{i,k}$ are all very small; otherwise one can assume $S$ equal to zero.  
The concentration $c_i$ of antibody $i$ is assumed to have, in absence of external antigens, only two values, conventionally $0$ or $1$ (in the presence of antigen concentrations $c_i$ might become $\geq 1$). The immune system state is determined by the values of all $c_i$'s for all possible antibodies $(i = 1;\dots;N)$. $h_i$ represents the total stimulatory/inhibitory (depending on its sign) effect of the whole network on the $i$-th antibody. $h_i$ is positive when the excitatory effect of the other antibodies is greater than the suppressive effect and then $c_i$ is one. Otherwise $h_i$ is negative and $c_i$ is zero.


\section{Methodology}
\label{sec:methodology}


The aim of our methodology is to extract local and global information from data applying TDA. It is suitable for studying the class of complex systems described in Section~\ref{sec:intro}. In particular, the output of the methodology is a set of models set up within the $S[B]$ paradigm~\cite{merelli2013non,merelli2014topology,MPT15}. In the $S[B]$ paradigm a model is specified using two levels of description, namely the $S$ \emph{global} or \emph{structural} level and the $B$ \emph{local} or \emph{behavioural} level, which are entangled in order to express the behavior of the system as a whole. The $S$ level describes how the system evolves following global information coming from the environment in which it is operating and from the interactions and the evolutions of the model entities of the $B$ level. 
Note that this approach has some similarities with hierarchical models like \textit{Hierarchical Automata} proposed by Mikk~et~al. in~\cite{mikk1997}. However, in our case the levels can not be seen as just different levels of abstraction, but they can exist only in the entangled $S[B]$ version. 

\subsection{From Data to Weighted Graphs}

We start from data that comes from observations, over time,  of the complex system under study. The first step to be accomplished is to represent such data as weighted graphs where:
\begin{itemize}
\item nodes represent the interacting components;
\item an arch, equipped with a weight, expresses an interaction (or a distance) between two components.
\end{itemize} 
Note that this is always possible; for instance the weight can be defined using statistical descriptors (correlation coefficients, scoring systems, and so on), metric functions (normalized Euclidean distance, Hamming distance, and so on) or domain dependent features.

Graphs are strictly related to topological spaces; in particular a graph can be thought as the skeleton of a topological space. Let us now introduce the description of a topological space that will be useful for our purposes. 

\begin{definition}{(Topological Space)}\\
\label{def:topo}
A topological space $\mathbb{X}$ can be described by a possibly infinite set of pairs of the form $(\beta_n,V_n)$, $n \geq 0$, where:
\begin{itemize}
\item $\beta_n$ is the $n$-th Betti number
\item $V_n \subseteq H_n$ is the set of generators of the homological class $H_n$, represented using an appropriate notation. 
\end{itemize}
\end{definition}

Concerning the notation for $V_n$, in case of $V_0$ we will use natural numbers, e.g.\ $\{0,1,2,3\}$, to identify the $0$-simplicies and in case of $V_1$ we will use a geometrical representation of the simplicial complex through an edge list, e.g.\ $[0,1]+[1,2]+[2,3]+[3,0]$. In case of $V_2$, a tetrahedron is the composition of triangles, e.g.\ $[0,1,2]+[0,1,3]+[1,2,3]+[0,2,3]$. In the rest of the paper, we use the notation $v \Subset v'$ for indicating that $v$ is a simplex of dimension strictly less than the dimension of the simplex $v'$ and it is a part of $v'$. For instance, in the example above $2 \Subset [1,2]$, $[0,1] \Subset [0,1,3]$ and $0 \Subset [0,1,3]$. For a more formal definition we refer to \cite{ghrist2008barcodes}.

\subsection{Form Weighted Graphs to Simplicial Complexes}
\label{sec:simpligraph}

Given a directed or undirected graph, it is possible to construct from it a simplicial complex following several approaches~\cite{jonsson2008simplicial}. In this methodology we apply the \textit{clique weight rank persistent homology} (CWRPH)~\cite{petri2013topological}. This innovative technique is based on the concept of a flag complex: given a graph $G$, the simplices of a clique complex $C(G)$ are the complete subgraphs of $G$ and the 0-simplices of $C(G)$ are the vertices of the subgraphs, (i.e., the complete subgraph complex). The maximal simplices are given by the collection of the vertices that make up the cliques of $G$. In the literature, a clique complex is also referred to as a flag complex. CWRPH describes a formal procedure for building a $C(G)$ from a weighted graph.

\textit{jHoles} is the first Java high-performance implementation of the CWRPH algorithm~\cite{binchi2014jholes}. It implements the following standard clique weight rank persistent homology: 
\begin{itemize}
\item[1] extract the descending (or ascending) list $W$ of all weights $w_t$ indexed by the discrete filter parameter $t$;
\item[2] list all maximal cliques of each connected component in $G$;
\item[3] find all the sub-cliques of each maximal clique; 
\item[4] for each sub-clique and maximal clique, rank it according to the index $t$ of the minimum (maximum) weight; note that these are clique simplicial complexes equipped with a filter value;

\end{itemize}

\begin{figure}
\begin{center}
\includegraphics[scale=0.70]{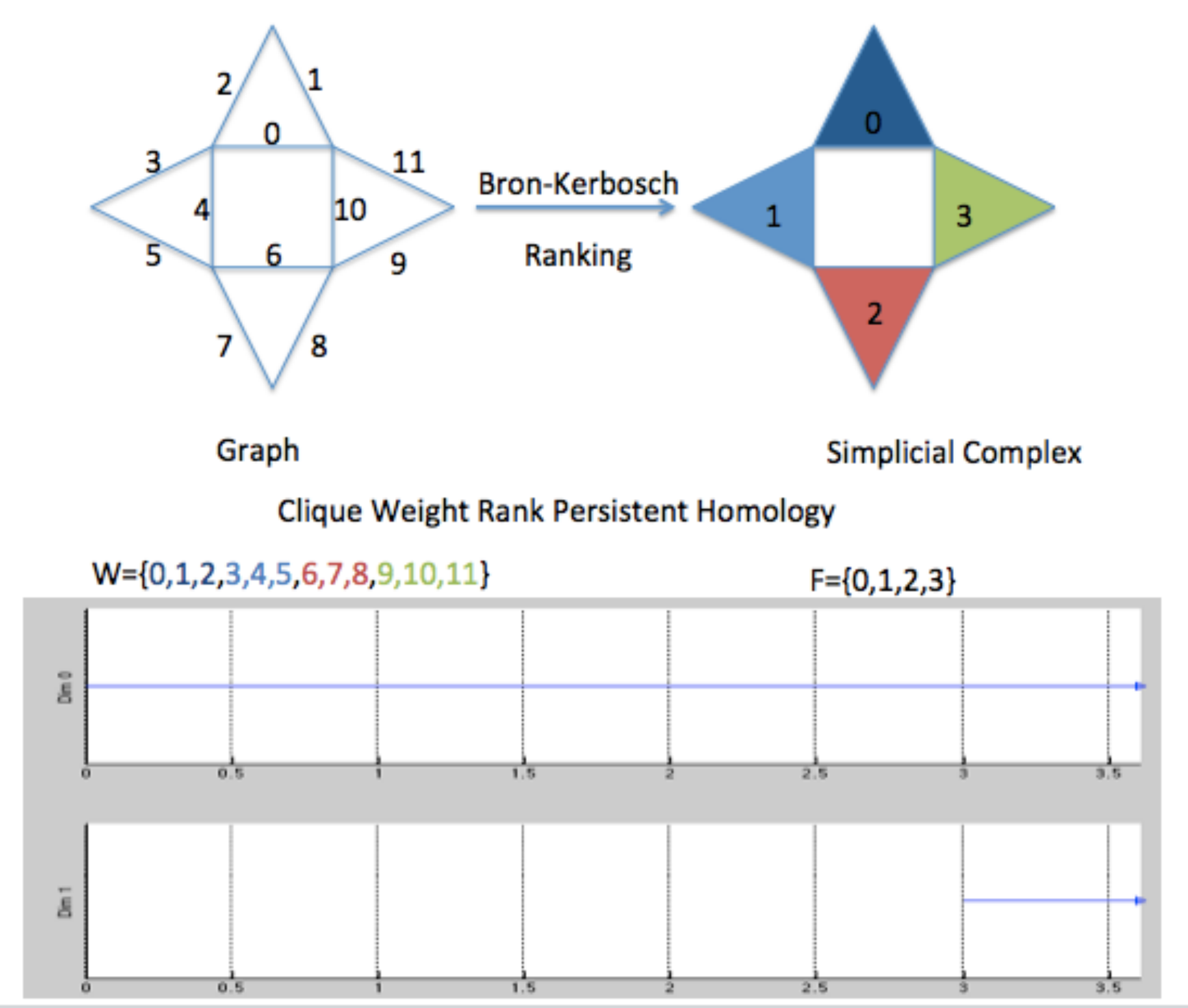}
\end{center}
\caption{jHoles application example. Construction of a topologicFlal space from an undirected weighted graph (up). The Betti barcode representing the evolution of topological invariants (down). Where W and F denote weights and filter values.}
\label{fig:jholes}
\end{figure}

In Figure~\ref{fig:jholes} it is shown an example of application of the CWRPH algorithm implemented in jHoles. Given the undirected weighted graph on the left, with weights $W$, jHoles finds four $3$-maximal cliques shown on the right with different colors; each of them is equivalent to a $2$-simplex (filled triangle). 
 
\subsection{From Simplicial Complexes to Computational Agents: Local Interactions}
\label{sec:local}

The last step of the jHoles computation takes the clique simplicial complexes calculated in step 4 (see Section~\ref{sec:simpligraph}) and computes the persistent homology giving as output Betti barcodes, intervals and generators.

Consider again Figure~\ref{fig:jholes}. The simplices are labeled with a filter value that is used as index during the computation of persistent homology~\cite{tausz2011javaplex}. Roughly speaking, the algorithm builds the topological space by including the simplices in the order given by their appearance filter value. In the example, the algorithm runs 4 iterations; at each step it introduces a simplex (giving filter values $F = \{0, 1, 2, 3\}$) that eventually is connected to the simplices previously introduced. Moreover, it computes the topological invariants (Betti numbers) of the newly constructed topological space. The pairs formed by Betti numbers and filter values are graphically represented with a Betti barcode. In this example the Betti numbers are $\beta_0 = 1$ meaning that there is only one connected component and $\beta_1=1$, indicating that a $1$-dimensional persistent topological hole is present. This hole is formed by four $1$-simplices and four $0$-simplices, which are its generators. The connected component appears at filter value $0$ in correspondence with the first $2$-simplex indexed $0$ and persists throughout the whole filtering, as indicated in the \textsc{Dim 0} of barcode in Figure~\ref{fig:jholes}. The persistent hole appears after that the four $2$-simplices are connected together at filter value $3$ (\textsc{Dim 1} of barcode). Note that in this example all the lines in the barcodes are persistent. However, this is not the case in general, because topological noise can be present (see Section~\ref{sec:introcs}).

At this point of our methodology, we use the generators of persistent homological classes of the derived topological space. In our point of view the $0$-simplices in the generators represent the more relevant interacting components of the complex system under study. They interact with other (homogeneous or heterogenous) entities or with the environment in which the system is immersed. The $n$-simplicies in the generators then tell how these components are connected together in the system. This is the information that can be automatically derived with the proposed methodology. 

It is very natural, in this view, to use the concept of \emph{Computational Agent} \cite{jennings2001agent,boccara2004modeling} to represent the identified more relevant interacting components. We want to derive a computational model for studying their behavior. An agent can execute a generic number of actions, possibly concurrently. However, TDA extracts information about connectivity among agents, but it does not provide any information regarding the actions executed by the entities. Thus, for obtaining a complete operational model of the interactions there is still the need of using domain-based knowledge. For instance, in Section~\ref{sec:casestudy} we will use the Jerne model described in Section~\ref{sec:introcs}.
 
Among other computational models available for expressing concurrent computations, in this case the \emph{true concurrent} characteristic of the kind of systems considered (not CPU-based), suggests the use of Higher Dimensional Automata for modeling the behavior of the agents. 

\subsubsection{Higher Dimensional Automata}

The problem of constructing a convenient model of computation that takes into account both the aspect of true concurrency, or non-interleaving concurrency, has been solved by using a geometrical approach. Several studies regarding a geometrical description of concurrency and automata have been published~\cite{mazurkiewicz1977concurrent,bernstein1987concurrency,lakshmivarahan1993symmetry}.

In 1991 Pratt published the formal definition of a Higher Dimensional Automaton based on the concept of $n$-complexes~\cite{pratt1991modeling}. This model is a generalization of automata to allow them to express non-interleaving concurrency. Pratt generalized the standard Finite State Automaton definition, based on states and edges, to general \textit{$n$-dimensional} objects where $n$ is any finite dimension. The basic idea is that an $n$-dimensional object stands for a $n$-dimensional transition representing the concurrent execution of $n$ actions.

\begin{figure}
\begin{center}
\includegraphics[scale=0.5]{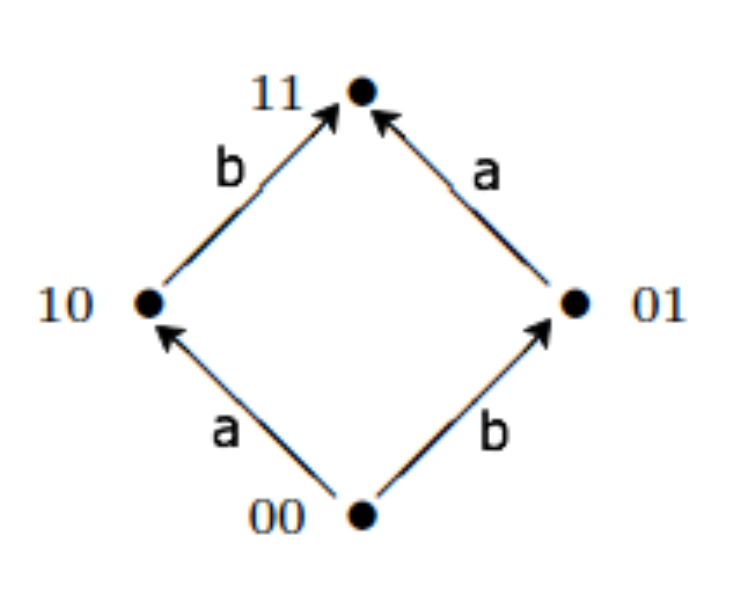}
\ 
\includegraphics[scale=0.5]{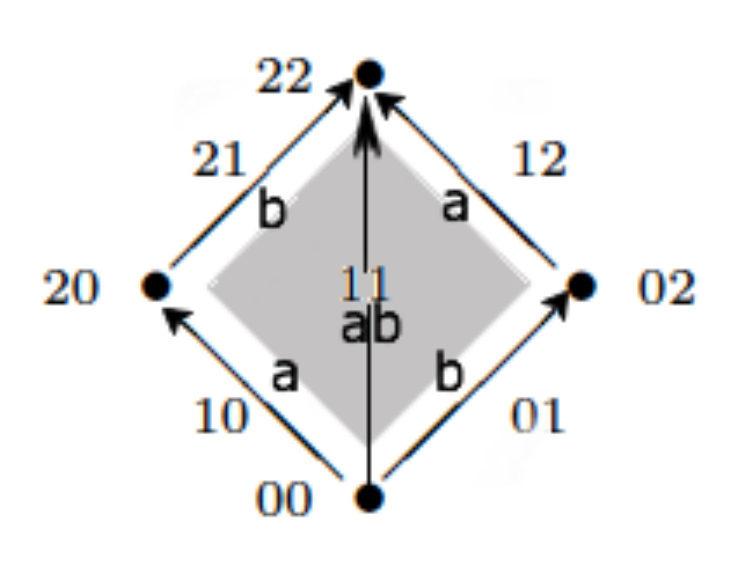}
\end{center}
\caption{A classical Finite State Automaton representing the interleaving of actions $a$ and $b$ (left). The corresponding Higher Dimentional Automaton (right).}
\label{fig:squares}
\end{figure}

A concept to manage with care is that of a ``hole''. Suppose to have the automaton on the left part of Figure~\ref{fig:squares}. The initial state $00$ has two outgoing edges labeled $a$ and $b$, respectively, followed by $b$ and $a$. Under the interleaving semantics, the automaton accepts both the words $ab$ and $ba$. However, such automaton contains a hole, indeed the interior of the square. Pratt suggested to replace the interleaving semantics of this automaton, towards true concurrency, by \textit{filling the hole}. The new surface, depicted on the right part of Figure~\ref{fig:squares}, represents the simultaneous execution of $a$ and $b$ without imposing any order. Conversely, the ``sculpting'' of the inner surface by proper transformations allows to represent with HDAs also classical \textit{non-determinism}. 

The initial formal definition of HDAs by Pratt can be found in \cite{pratt1991modeling}. However, in the subsequent papers \cite{gupta1994chu, pratt2000higher}, a more intuitive and concise definition was given by using Chu spaces. For the sake of simplicity, we recall here this definition. A Chu space \cite{pratt2000higher} is a structure over an alphabet $\Sigma$, namely a rectangular array whose entries are drawn from $\Sigma$. $\Sigma=\{0,1\}$ is appropriate for representing ordinary event structures \cite{winskel1982event}, whose events may be either \emph{unstarted} or \emph{finished}. The alphabet can be extended to $\Sigma=\{0,1,2\}$ meaning that an action can be \textit{unstarted}, \textit{executing} or \textit{finished}. The 3-length alphabet is used for HDAs. 

	\begin{definition}{(Chu Space)}\\

A Chu space $\mathbb{A}$ is a tuple $(A,~r,~X)$ over an alphabet $\Sigma$ where:

		\begin{itemize}
\item $A$ is a set of \emph{actions} or \emph{events};
\item $X$ is a set of \emph{states};
\item $r \colon (A \times X ) \rightarrow \Sigma$ is a function representing whether or not an event belongs to a state.
		\end{itemize}
	\end{definition}
		
Consider the FSA on the left part of Figure~\ref{fig:squares}. Its representation as a Chu space with $\Sigma = \{0,1\}$ can be given with the following matrix representing the function $r$ of the Chu space:

\begin{center}
\begin{tabular}{|l|l|l|l|l|}
\hline
a & 0 & 0 & 1 & 1\\
\hline
b & 0 & 1& 0&1\\
\hline
\end{tabular}
\end{center}

Every column represents a possible state of the automaton. For instance, state $00$ is the initial one where both $a$ and $b$ are unstarted, state $01$ is the one in which $a$ is still unstarted and $b$ is finished, and so on. 

Consider now the HDA on the right part of Figure~\ref{fig:squares}. Its representation as a Chu space with $\Sigma = \{0,1,2\}$ can be given as follows:

\begin{center}
\begin{tabular}{|l|l|l|l|l|l|l|l|l|l|}
\hline
a & 0 & 0 & 0 & 1 & 2 & 1 & 1 & 2 & 2\\
\hline
b & 0 & 1 & 2 & 0 & 0 & 1 & 2 & 1 & 2\\
\hline
\end{tabular}
\end{center}

State $01$ now is the one in which action $a$ is unstarted and action $b$ is executing, state $12$ is the one in which action $a$ is executing and action $b$ is finished, and so on. 
 
Chu spaces have also been used for the study of concurrent programs. In~\cite{du2010modeling} Du et al.\ proposed an enriched process algebra for the Chu spaces for studying the concurrency of object-oriented programming languages. In~\cite{ivanov2008modeling, ivanov2008automatic} Ivanov presented an algorithm for generating Chu space models for describing the behaviors of complex non-iterated systems, including $n$-ary dependencies.  

\subsubsection{Derivation of HDAs from Topological Generators}

The next step of the methodology is to derive from the simplicial complex constructed in Section~\ref{sec:simpligraph} an HDA representing the behavioral level of the model within the $S[B]$ paradigm. Such HDA is the parallel composition of the HDAs representing the behaviour of each computational agent identified by the $0$-simplices in the generators $V_0$. For the HDA we will use the Chu space representation. It is useful for our purposes to define the set $A$ of the Chu space representing the HDA in such a way that the name of the actions carry additional significant information. 

Let $\mathbb{X}$ be representation of a topological space as in Definition~\ref{def:topo}, i.e.\ pairs of Betti numbers $\beta_n$ and the corresponding generators $V_n$ of the homological classes.

\begin{definition}{(Labels of Actions)}\\
The labels $A$ of the Chu space representing the HDA are of the form $(\mathrm{action}, \mathrm{source},\mathrm{target})$ where:
\begin{itemize}
\item $\mathrm{action}$ is a description name;
\item $\mathrm{source}$ and $\mathrm{target}$ are the identifiers of two $0$-simplices $v,v' \in V_0$ and there exists $v_n \in V_n$ ($n>0$) such that $v \Subset v_n$ and $v' \Subset v_n$.
\end{itemize}
\end{definition}

Note that in general, when no particular constraints are present in the domain knowledge, the actions are bi-directional, i.e.\ there is an action $a_0$ from $v$ to $v'$ and the same action $a_0$ from $v'$ to $v$. Moreover, the domain knowledge should suggest a certain number of actions and assign to them a meaningful name.  

\subsection{From Simplicial Complexes to Persistent Entropy: Global Information}
\label{sec:persistent-entropy}

Let us consider again the Betti barcodes, intervals and generators calculated from simplicial complexes as described at the beginning of Section~\ref{sec:local}. In order to extract global information from the data-derived topological space, the next step of the methodology prescribes to compute a newly introduced notion that we call \textit{persistent entropy}. This entropy measure is basically calculated using the persistent Betti barcodes and the topological noise of the underlying topological space. By definition, the value of the persistent entropy is strongly related to the topological structures derived from data. 

Diaz et al.\ defined an entropy based on the persistent barcode (Definition 3 of \cite{chintakunta2015entropy}). The aim of their paper is the definition of an entropy-driven algorithm for finding the best filtration of a set of simplices. We argue that when the filtration is given their entropy can be easily extended without loosing the interpretation $\grave{a}$ la Shannon. Here we propose to use the maximum of the filtration value plus one as upper bound of a persistent barcode.

\begin{definition}{(Persistent Entropy)}
\label{def:persistent-entropy}
Given a filtered topological space, let $F$ be the set of its filtration values and let $J = \{0, 1, \ldots, $n$\}$ be a set of indices for the lines appearing in the whole barcode, independently from the dimensions. A line corresponding to topological noise 
is denoted by $[a_j~;~b_j]$, with $j \in J$. Instead of $[a_j~;~\infty)$, a persistent topological feature is denoted by $[a_j~;b_j = m)$ where we set $m = \max\{F\} + 1$.

The \emph{Persistent Entropy} $H$ of the topological space is calculated as follows: 
$$
H=-\sum_{j \in J} p_j  log(p_j)
$$
where $p_j=\frac{l_j}{L}$, $l_j=b_j - a_j$, and $L=\sum_{j\in J}l_j$.
\end{definition}

Note that the maximum persistent entropy corresponds to the situation in which all the lines in the barcode are of equal length. Conversely, the value of the persistent entropy decreases as more lines of different lenght are present. 

Consider again the topological space derived in Figure~\ref{fig:jholes}. The set $F={0,1,2,3}$, let $J = \{0,1\}$ where $0$ is the index of the line in \textsc{Dim 0} and $1$ is the line in \textsc{Dim 1}, both persistent in this case. Then, $m = 4$ and the lines are $[0,4)$ and $[3,4)$ yielding $H = 0.5$. 

\subsection{The Derived $S[B]$ Model}

In the vision of the $S[B]$ paradigm global information constraints local interactions within a certain subspace of all the possible interactions according to equilibrium conditions (steady states) characterizing the system. Whenever external or internal factors alter this equilibrium, the system reacts by entering an ``adaptation'' phase towards a new equilibrium~\cite{MPT15}. Thus, if we observe the system at any moment, like we do when we take samples of the data, it will be either in a stable condition or in an adapting phase because the global information (strictly bond to the current local configuration) is violating the current equilibrium condition. As the system evolves, it can go back to the previous steady state or it may eventually reach a new kind of known or unknown equilibrium.  

In our methodology persistent entropy is the global information that, at the $S$ level, naturally constraints the local interactions at the $B$ level represented by the HDA constructed as described in Section~\ref{sec:local}. Note that we derive from data and from domain specific knowledge a set of models each of which give an $S[B]$ representation of the system at the moment of the observation. The extraction of the rules governing the evolution of the system between one observation and the subsequent one is beyond the scope of this work.

\subsection{Global Evolution of the Extracted Models}

The evolution over time of the persistent entropy can be used for detecting when the system reacts to stimuli, i.e., starts an adaptation phase. Let us consider the chronogram plotting the values of the persistent entropy along the observed sequence of data of the system. Indeed, a peak in the chart means an abrupt change in the topology of the simplicial complexes. This information may be used to find the intervals \textit{before, during and after} a stimuli (or a phase transition)~\cite{han2012graph}. On the other hand, a plateau in the chart corresponds to an equilibrium condition, namely a steady state. The relevant regions can then be used for deriving a state machine representing the global observed evolution of the system.

\textit{Automata} \cite{hopcroft1979introduction} are one of the mostly used formalism for modeling systems (hardware, software, physical systems, and so on). They are a very familiar and useful model that extends the concept of directed graph associating to the nodes and the edges the meaning of \textit{states} and \textit{transitions}, respectively. A state describes a static feature of the behavior of the system during its evolution, while a transition models how the system can evolve from one state to another. In the literature, several types of automata have been discussed~\cite{park1981concurrency,alur1994theory,henzinger2000theory,rabin1963probabilistic}. Among others, a generalization of the formalism of automata and, more generally, transition systems is that of \textit{Uniform Labeled Transition Systems} (ULTraS)~\cite{bernardo2013uniform,BT13}. In this paper our automata are equipped with topological information coming from data. 

Following the approach of \cite{MPT15}, we use a notion of \emph{observable} that is derived from the behavioral level $B$ and is used in $S$ to define constraints associated to the $S$ states. Such constraints identify the known \emph{steady states} of the $S[B]$ system and, when violated, put the whole system in an \emph{adapting} phase.

\begin{definition}{(Persistent Entropy Automaton)}\\
A Persistent Entropy Automaton (PEA) is a tuple $S=(R, \Lambda, r_0, H, \goes{}_S, L)$ where:
\begin{itemize}
\item $R$ is a set of steady states;
\item $\Lambda$ is a set of labels;
\item $r_0 \in R$ is the initial steady state;
\item $H$ is the observable variable, corresponding to the current value of Persistent Entropy;
\item $\goes{}_S \subseteq R \times \Lambda \times R$ is a labelled transitions relation;
\item $L \colon R \rightarrow \Phi_H$ is a labeling function associating to each state $r \in R$ an \emph{invariant condition} on $H$ representing the global constraints that characterize the equilibrium configuration expressed by $r$.
\end{itemize}
\end{definition}

\section{$S[B]$ Model of the Case Study}
\label{sec:casestudy}

Accordingly to Jerne~\cite{jerne1974towards}, the Idiotypic Network (IN) can be described by three main states: \textit{virgin, activation} and \textit{memory}. In the \textit{virgin} state there are not antibodies but only non-specific T-cells that start their activities after the activation signal sent by B-cells. B-cells recognize that the environment has been disrupted by pathogens. Then, the IN \textit{proliferates} antibodies and reach the \textit{activation} state, which is not a steady state. After the activation, the IN performs the \textit{immunization} during which the antibodies play a dual role. In fact, an antibody can be seen both as a \textit{self} protein (a protein of the organism) or a \textit{non-self} protein (a pathogen to be suppressed). After the \textit{immunization} action the IN reaches the \textit{memory} state. This state represents a steady condition in which there is only a selection of antibodies. All the \textit{transitions and states} of IN are characterized by the fact that antibodies perform basically two actions: \textit{elicitation} and \textit{suppression}.    

In the rest of this section we report about the application of our methodology to the IN simulated with \textit{C-ImmSim}~\cite{bernaschi2001design,castiglione2015,mancini2012}. We executed several (in the order of hundreds) simulations, each of them characterized by:
\begin{itemize}
\item a lifespan of 2190 ticks, where a tick corresponds to three days;
\item a repertoire of at most $10^{12}$ antibodies, i.e., the maximum number of antibodies available during the whole simulation;
\item an antigen volume $V = 10 \mu L$. 
\end{itemize}

During the simulation an antigen is injected and, after an unknown (simulated) transient period of time, the same antigen is injected again. 

\subsection{From data to weighted graph}

In \textit{C-ImmSim} each idiotype (both antigens and antibodies) is represented with a bit-string, in our case of 12 bit length. Two idiotypes $A_j$ and $A_k$ interact if and only if their Hamming distance $d(A_j, A_k)$ is such that $11\leq d(A_j, A_k) \leq 12$. The pair-wise distances among all the idiotypes are stored in a matrix $J$, the so-called \textit{Affinity matrix}. 

However, the so formed affinity matrix is not very realistic because it does not take into account the volume of the idiotypes. For this reason, in this case study we decided to replaced it with a \textit{Coexistence Matrix} $C$ where each element is a coexistent index.

\begin{definition}{(Coexistence Index)}\\
Given the Hamming distance $d(Ab_j(t),Ab_k(t))$ between two antibodies and their volumes $[Ab_{j}(t)], [Ab_{k}(t)]$ at tick $t$, their 
\textit{coexistence index} is defined as follows:
\begin{equation}
C_{Ab_{j,k}}(t)=\frac{d(Ab_j(t),Ab_k(t)) \cdot [Ab_{j}(t)]\cdot[Ab_{k}(t)]}{\sum_{l=1}^{n} [Ab_{l}(t)]} 
\label{eq:eq1}
\end{equation}
\end{definition}

Equation~\ref{eq:eq1} expresses the fact that for lower values of affinity the volume must be more significant because the match between antibodies is less probable.

\begin{figure}[ht!]
\centering{
\includegraphics[width=\textwidth]{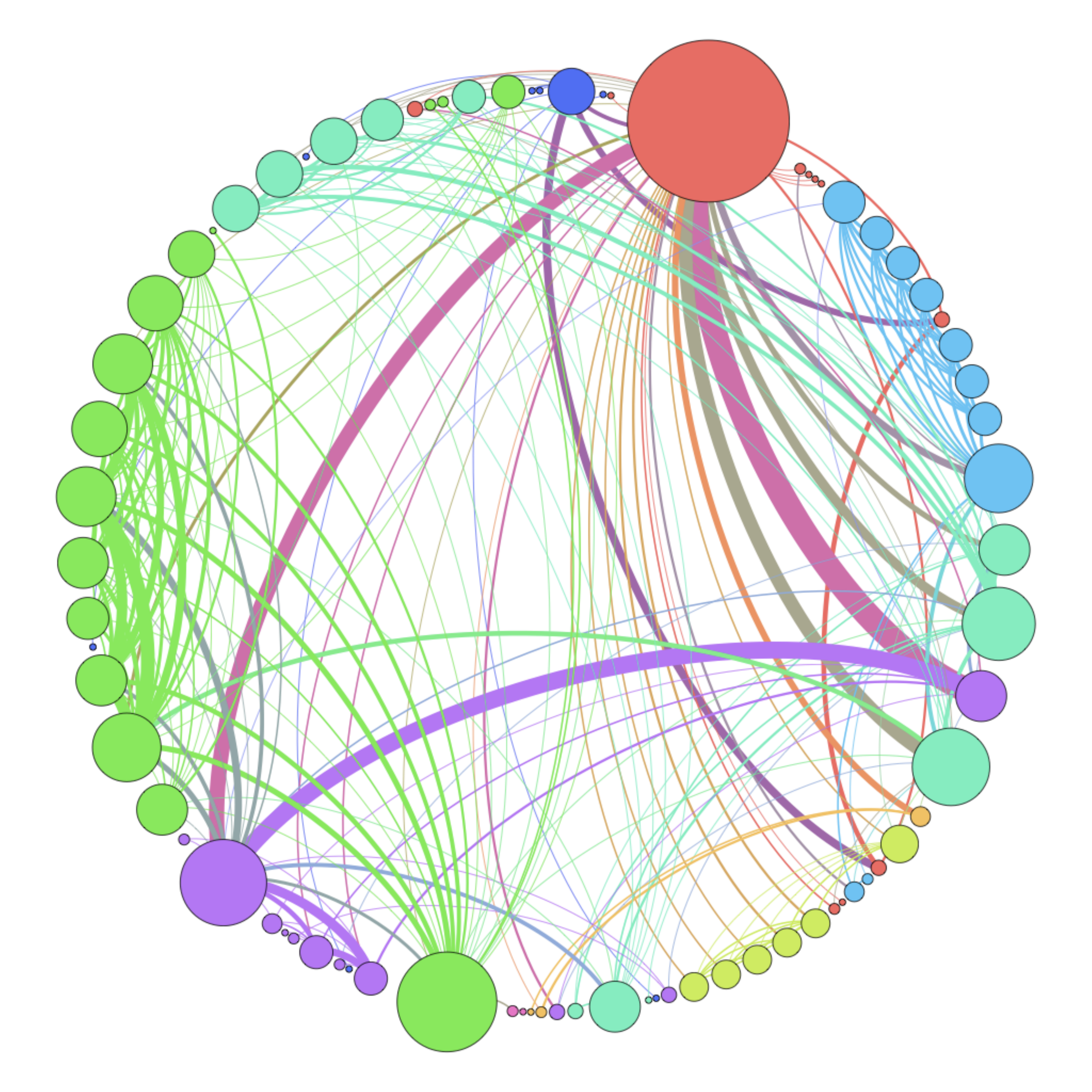}
\caption{Example of immune network at the end of the simulation. The thickness of the arcs is proportional to their weight, the diameter of nodes is proportional to the number of incident edges.}
\label{fig:idiotypes}
}
\end{figure}

The coexistence matrix $C$ is a symmetric matrix and each element is taken to represent a weighted arc of an undirected graph. In this way we obtained a graph representation of our data taken from the simulation. Figure~\ref{fig:idiotypes} shows the weighted idiotypic network at the end of the simulation.

\subsection{From Weighted Graph to Simplicial Complex}

Ee computed the coexistence matrix (see Equation~\ref{eq:eq1}) and we used the weighted idiotypic network as the input for the persistent homology computation. As indicated by the methodology we used the CWRPH, recent development in TDA, providing a new approach to the study of weighted networks. One of the main advantages of this approach is that it preserves the complete topological and weight information, allowing to focus on special mesoscopic structures, e.g., weighted network holes, which connect the network's weight-degree structure to its homological backbone~\cite{petri2013topological}.

We used jHoles~\cite{binchi2014jholes} to process our weighted graph towards the production of the simplicial complex as described in Section~\ref{sec:simpligraph}, points 1-4 of the jHoles algorithm.

\subsection{Computing Persistent Homology}
\label{sec:phom}
The persistent homology of the simplicial complex obtained above is then computed with jHoles (Section~\ref{sec:simpligraph}, point 5 of the jHoles algorithm). The output is a collection of persistent barcodes, one for each \textsc{Dim}, and the set of the generators for the persistent topological features. An example of the output is given in Figure~\ref{fig:generators}. This information is used for deriving the HDAs and for computing the persistent entropy.

\begin{figure}
\begin{center}
$\beta_0$:\\
$[0.0, infinity): [16]$\\
$\beta_1$:\\
$[7.0, infinity): [320,3775] + [256,3775] + [320,3839] + [256,3839]$
$[6.0, infinity): [256,3839] + [256,3711] + [384,3711] + [384,3839]$
$[8.0, infinity): [260,3835] + [260,3839] + [256,3835] + [256,3839]$
\end{center}
\caption{Example of generators.}
\label{fig:generators}
\end{figure}

\subsubsection{HDA derivation}
From the Jerne model (see Section~\ref{sec:introcs}) it is well known that antibodies exhibit a \textit{true-concurrent behavior}, i.e., they can perform elicitation and suppression actions simultaneously on different targets. Using this domain knowledge  we modeled the more relevant antibodies, given by the generators of the persistent topological features, with an HDA represented as a Chu space over the 3-length alphabet. A graphical representation of a Chu space is often given using an \textit{Hasse diagram}. Hasse diagram helps to easily deduce the set of admissible traces, dropping out automatically the \textit{forbidden states}. In these diagrams the computation proceeds in the upwards direction and represents an execution path in the automaton~\cite{gupta1994chu}. As an example, Figure~\ref{fig:latticeAb} shows the Hasse diagram representing the behaviour of a single antibody $Ab$ executing two possible actions concurrently, whose corresponding Chu space representation is given in Table~\ref{tab:Abchu}. Note that there is not a direct path between state $(1,0)$ to $(0,1)$ meaning that if action $(\mathrm{elicits}, Ab, \mathrm{target})$ is \textit{executing} and action $(\mathrm{reduces}, Ab, \mathrm{target})$ is \textit{unstarted}, then the first cannot go back to the unstarted situation while the second becomes executing (or vice-versa). Instead, from state $(1,0)$ it is possible to go to state $(1,1)$ meaning that both actions can be executing at the same time. Finally, if in the Hasse diagram the line between $(1,0)$ and $(1,1)$ was not present, that would mean that action $(\mathrm{reduces}, Ab, \mathrm{target})$ cannot start executing before action $(\mathrm{elicits}, Ab, \mathrm{target})$ has become \textit{finished}. 

\begin{table}
\begin{center}
   \begin{tabular}{|r | rrrrrrrrr|}
    \hline
    Ab & s1    & s2    & s3    & s4    & s5 & s6 & s7& s8& s9\\ 
    \hline
    $(\mathrm{elicits}, Ab, \mathrm{target})$ 	& 0  & 0 & 0&1  & 1 & 1& 2 & 2 & 2 \\
    $(\mathrm{reduces}, Ab, \mathrm{target})$  & 0  & 1 &2& 0  & 1 & 2& 0 & 1 & 2\\
    \hline 
    \end{tabular}
\end{center}
 \caption{Chu space representation of an HDA modeling a generic antibody $Ab$ performing two actions.}
 \label{tab:Abchu}
\end{table}

As a more concrete example, we report the HDA modeling the behavior of two coupled antibodies $Ab_1$ and $Ab_13$ resulting from the simulations. Figure~\ref{fig:grossa} shows the matrix of the Chu space representing such subsystem. Figure~\ref{fig:latticeAbAb} shows the relative Hasse diagram.

\begin{figure}

\begin{center}
    \begin{tikzpicture}[scale=.5]
    \draw[fill] (0,0) circle (.05cm) node[below] {$(0,0)$};
    \draw (0,0) -- (-2,2);
    \draw (0,0) -- (2,2);

    \draw[fill] (-2,2) circle (.05cm) node[left] {$(1,0)$};
   \draw (-2,2) -- (0,4);
   \draw (-2,2) -- (-4,4);

    \draw[fill] (0,4) circle (.05cm) node[below] {$(1,1)$};
   \draw (0,4) -- (-2,6);
    \draw (0,4) -- (2,6);
    
     \draw[fill] (2,2) circle (.05cm) node[right] {$(0,1)$};
   \draw (2,2) -- (0,4);
   \draw (2,2) -- (4,4);
    
    \draw[fill] (-4,4) circle (.05cm) node[left] {$(2,0)$};
    \draw (-4,4) -- (-2,6);

     \draw[fill] (4,4) circle (.05cm) node[right] {$(0,2)$};
    \draw (4,4) -- (2,6);
   
    \draw[fill] (-2,6) circle (.05cm) node[left] {$(2,1)$};
   \draw (-2,6) -- (0,8);
   
    \draw[fill] (2,6) circle (.05cm) node[right] {$(1,2)$};
   \draw (2,6) -- (0,8);
    
    \draw[fill] (0,8) circle (.05cm) node[above] {$(2,2)$};

    \end{tikzpicture}
    
\end{center}

\caption{Hasse diagram for Chu space representing a generic antibody $Ab_i$}
\label{fig:latticeAb}
\end{figure}
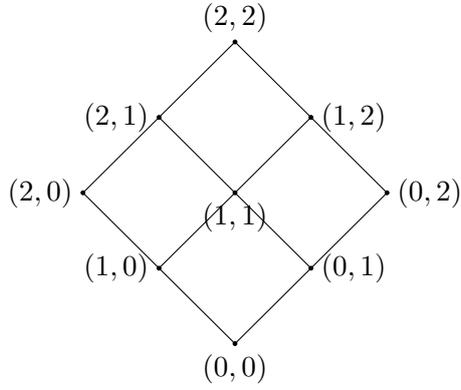

\begin{figure}
\noindent\par
\noindent\makebox[\textwidth][l]{%

 \begin{minipage}{15cm}
 \centering{ $\{Ab_1 \mid Ab_{13}\}$ }
  \[ \left|
  \begin{subequations}
   \begin{array}{l | lllllllllllllllll}
    \toprule
	Ab_1| Ab_{13} & s1 & s2 & s3 & s4 & s5 & s6 & s7 & s8 & s9 & s10 & s11 & s12 & s13 & s14 & s15 & s16 & s17  \\ \hline
	 \midrule
	elicitsAb_{i,j} & 0 & 1 & 2 & 0 & 0 & 0 & 0 & 0 & 0 & 1 & 1 & 1 & 1 & 2 & 2 & 2 & 2 \\
	 \midrule
	reducesAb_{i,j} & 0 & 0 & 0 & 1 & 2 & 0 & 0 & 0 & 0 & 0 & 0 & 0 & 0 & 0 & 0 & 0 & 0\\ 
	 \midrule
	elicitsAb_{j,i} & 0 & 0 & 0 & 0 & 0 & 1 & 2 & 0 & 0 & 1 & 0 & 2 & 0 & 1 & 0 & 2 & 0 \\ 
	 \midrule
	reducesAb_{j,i} & 0 & 0 & 0 & 0 & 0 & 0 & 0 & 1 & 2 & 0 & 1 & 0 & 2 & 0 & 1 & 0 & 2  \\\bottomrule 
    \end{array}
    \end{subequations}
     \right |\] 
     \label{sub:chuabab1}
    
    \end{minipage}}
       \noindent
\noindent\makebox[\textwidth][l]{%
 \begin{minipage}{10cm}
     \[ \left|
  \begin{subequations}
     \begin{array}{l | llllllll}
    \toprule
	Ab_1| Ab_{13} & s18 & s19 & s20 & s21 & s22 & s23 & s24 & s25 \\ \hline
	 \midrule
	elicitsAb_{i,j}  & 0 & 0 & 0 & 0 & 0 & 0 & 0 & 0 \\
	 \midrule
	reducesAb_{i,j} &  1 & 1 & 1 & 1 & 2 & 2 & 2 & 2 \\ 
	 \midrule
	elicitsAb_{j,i} & 1 & 0 & 2 & 0 & 1 & 0 & 2 & 0 \\ 
	 \midrule
	reducesAb_{j,i} & 0 & 1 & 0 & 2 & 0 & 1 & 0 & 2 \\\bottomrule 
    \end{array}
    \end{subequations}
     \right |\] 
     \label{sub:chuabab2}
       \par
    \end{minipage}}
\caption{Chu space representing the subsystem $\{Ab_{1} \mid Ab_{13}\}$.}
\label{fig:grossa}
\end{figure}
 
\begin{sidewaysfigure}
\begin{center}

\begin{tikzpicture}[scale=.8]
  \node (zero) at (0,-2) {$\emptyset$};

  \node (a) at (-6,0) {$(1,0,0,0)$};
  \node (b) at (-2,0) {$(0,1,0,0)$};
  \node (c) at (2,0) {$(0,0,1,0)$};
  \node (d) at (6,0) {$(0,0,0,1)$};
  \node (e) at (-12,3) {$(2,0,0,0)$};
  \node (f) at (-2,3) {$(0,2,0,0)$};
  \node (g) at (2,3) {$(0,0,2,0)$};
  \node (h) at (12,3) {$(0,0,0,2)$};
  \node (ac) at (-7.5,1.5) {$(1,0,1,0)$};
  \node (bc) at (-4.5,1.5) {$(0,1,1,0)$};
  \node (bd) at (4.5,1.5) {$(0,1,0,1)$};
  \node (ad) at (7.5,1.5) {$(1,0,0,1)$};
   \node (eac) at (-10,6) {$(2,0,1,0)$};
  \node (gac) at (-2,6) {$(1,0,2,0)$};
  \node (fbc) at (2,6) {$(0,2,1,0)$};
  \node (gbc) at (10,6) {$(0,1,2,0)$};
  \node (ead) at (-7.5,6) {$(2,0,0,1)$};
  \node (had) at (-4.5,6) {$(1,0,0,2)$};
  \node (fbd) at (4.5,6) {$(0,2,0,1)$};
  \node (bdh) at (7.5,6) {$(0,1,0,2)$};
\node (eacgac) at (-6,9) {$(2,0,2,0)$};
  \node (eadhad) at (-2,9) {$(2,0,0,2)$};
  \node (fbcgbc) at (2,9) {$(0,2,2,0)$};
  \node (fbdbdh) at (6,9) {$(0,2,0,2)$};
\draw (eac) -- (eacgac);
\draw (gac) -- (eacgac);

\draw (fbc) -- (fbcgbc);
\draw (gbc) -- (fbcgbc);

\draw (ead) -- (eadhad);
\draw (had) -- (eadhad);

\draw (fbd) -- (fbdbdh);
\draw (bdh) -- (fbdbdh);

\draw (f) -- (fbc);
\draw (bc) -- (fbc);

\draw (e) -- (eac);
\draw (ac) -- (eac);

\draw (g) -- (gac);
\draw (ac) -- (gac);

\draw (g) -- (gbc);
\draw (bc) -- (gbc);

\draw (e) -- (ead);
\draw (ad) -- (ead);

\draw (h) -- (had);
\draw (ad) -- (had);

\draw (f) -- (fbd);
\draw (bd) -- (fbd);

\draw (b) -- (bdh);
\draw (bd) -- (bdh);

\draw (a) -- (ad);
\draw (d) -- (ad);
  
\draw (a) -- (ac);
\draw (c) -- (ac);
  
\draw (b) -- (bc);
\draw (c) -- (bc);
  
\draw (b) -- (bd);
\draw (d) -- (bd);

\draw (zero) -- (a) -- (e);
\draw (zero) -- (b) -- (f);
\draw (zero) -- (c) -- (g);
\draw (zero) -- (d) -- (h) ;
 
  \end{tikzpicture}

\end{center}
\caption{Hasse diagram for Chu space representing the subsystem $\{Ab_{1} \mid Ab_{13}\}$.}
\label{fig:latticeAbAb}
\end{sidewaysfigure}

\subsubsection{Persistent Entropy Calculation}

From the barcodes obtained in Section~\ref{sec:phom} we calculated, by using Definition~\ref{def:persistent-entropy}, the value $H(t)$ of persistent entropy associated to every observation at time $t$. Note that in this case it is possible to build a chart of $H$ over time because we know that the data are subsequent as they come from the same simulation. In general, for comparing two values of $H$ one should define the so-called \emph{mass function} \cite{stewart2009probability}.

\begin{figure}
\centering{
\includegraphics[width=0.8\textwidth]{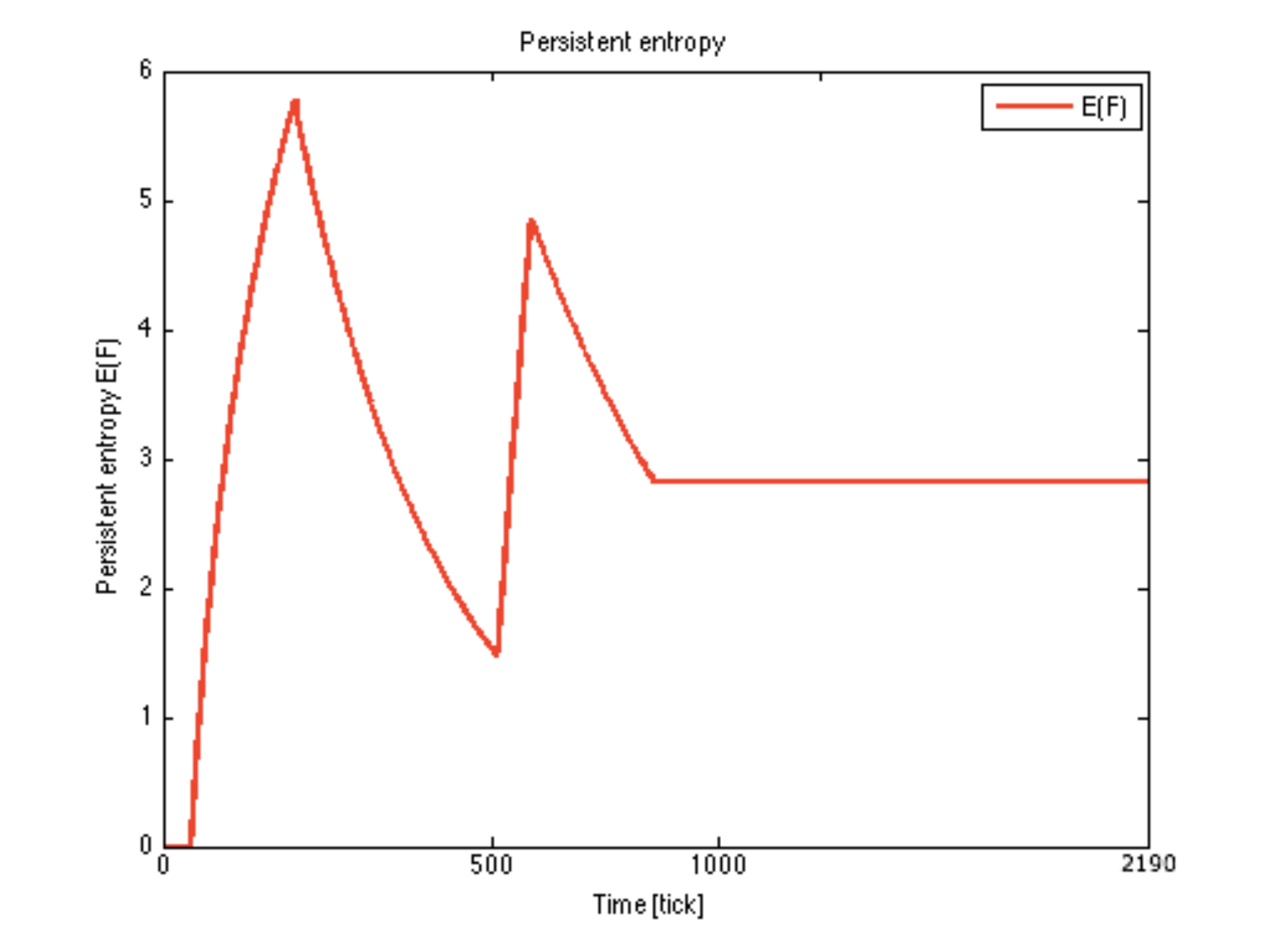}
\caption{Persistent entropy of immune system. The difference between the peaks amplitude is motivated by the fact that before the second peaks the antibodies have been already stimulated and the \textit{immune memory} has been reached, so the system is more reactive and is faster in the suppression of the antigen, thus the the entropy values for the steady state is H=2.87. }
\label{fig:entropy4}
}
\end{figure}

\subsection{Analysis of the Plot of Persistent Entropy versus Time}

Figure~\ref{fig:entropy4} show the charts of $H(t)$. 
Persistent entropy recognizes the dynamics of the immune system: a peak in the chart means \textit{immune activation} while a plateau represents the \textit{immune memory}. An upward curve between a peak and a plateau means \textit{proliferation} while a backward curve means \textit{immune response}. From the comparison of the charts it is possible to recognize that the system is stimulated twice (Figure~\ref{fig:entropy4}).  

Note that from the charts it is also possible to derive a discrete estimation of the first and second derivatives of $H(t)$ by using finite differences. These will be useful in the following section.

\subsection{PEA Derivation}
Analyzing the chart of persistent entropy we were able to recognize two steady states, namely \textit{virgin} and \textit{memory}. The \textit{virgin} state of IN is characterized by an invariant condition $H=0 \, \wedge \,\dot{H} = 0$. This is the initial state in the PEA that we derive. The second steady state corresponds to a plateau in the chart and it is characterize by an invariant condition $H>0 \, \wedge \, \dot{H} = 0$. 

Note that in the chart of PE (Figure \ref{fig:entropy4}) two peaks are present. This reflects the fact that the system was stimulated twice and, thus, entered an adaptation phase from state \textit{memory} to itself. This then suggests that a self-transition should be added to the state \textit{memory}. 

Finally, even if we do not observe it in our simulation, we know from domain specific knowledge that when in \textit{virgin} state, if the received stimulus is not grater than a certain threshold then the immune activation does not start at all and after a while the system goes back to the initial state again. This means that also in the initial state of the PEA a self-loop transition should be added.

\begin{figure}
\begin{center}

\tikzset{initial text={}}
\begin{tikzpicture}[->,>=stealth',shorten >=1pt,auto,node distance=2.8cm,
                    semithick]
\tikzstyle{every state}=[fill=white,draw=red,text=black]

      \node[state,initial] (v)   [align=center]{Virgin\\$H = 0,~\dot{H}=0$}; 
   \node[state](m) [ right=of v] [align=center]{Memory\\ $H >0,~\dot{H}= 0$};
    \path[->] 
    (v) edge  node {Immunization} (m)
    (v) edge [loop above] node {Resistance} (v)
    (m) edge [loop above] node  {Adaptability} (m);
    
    \end{tikzpicture}
   
\end{center}
\caption{PEA representing the S level of the idiotypic network.}
\label{fig:SB1}
\end{figure}
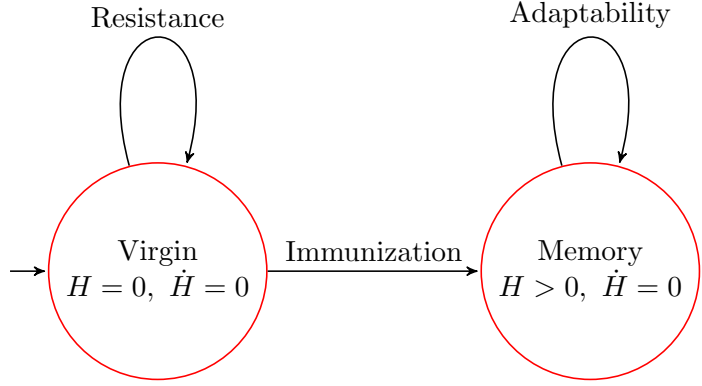

\section{Conclusions}
\label{sec:conclusions}

In this work we defined a methodology, based on Topological Data Analysis, for deriving a sequence of two-level models, within the $S[B]$ paradigm, of a complex system from data. In particular, each  data observation is initially represented with a weighted graph (network) and then its structure is used for building a richer topological space. Then, the topological space is characterized by computing persistent homology, from which local and global information about the complex system is extracted. The local information, about interacting computational agents, is represented by Higher Dimensional Automata. The global dynamics is studied by persistent entropy, which is newly introduced entropy measure linking together information theory and topology. We showed that the methodology can be effectively used by applying it to a case study about the idiotypic network of mammal immune system.
As future work we plan to investigate the possibility of using Higher Dimensional Automata also for representing the global dynamics. In this way, we can improve the scalability of our methodology because an $S[B]$ derived for a certain system can be used as the model of an agent in the study of a larger complex system of which it is an interactive component.

\section*{Acknowledgments}
We acknowledge the financial support of the Future and Emerging Technologies
(FET) programme within the Seventh Framework Programme (FP7) for Research
of the European Commission, under the FP7 FET-Proactive Call 8 - DyMCS, Grant
Agreement TOPDRIM, number FP7-ICT-318121. The authors thank Mario Rasetti for being their mentor unique and invaluable during the years of TOPDRIM project and especially for his helpful suggestions but also criticisms. \\

\end{document}